\documentclass[aps,prl,twocolumn,groupedaddress,preprintnumbers,nofootinbib]{revtex4}
\usepackage{graphicx}
\begin{document}

\newcommand{\lang}{\mathcal{L}}

\newcommand{\pd}[2]{\frac{\partial #1}{\partial #2}}

\newcommand{\half}{{\frac{1}{2}}}
\newcommand{\beq} {\begin{eqnarray}}
\newcommand{\eeq} {\end{eqnarray}}
\newcommand{\nn}{\nonumber}
\def\ltap{\ \raise.3ex\hbox{$<$\kern-.75em\lower1ex\hbox{$\sim$}}\ }
\def\gtap{\ \raise.3ex\hbox{$>$\kern-.75em\lower1ex\hbox{$\sim$}}\ }
\newcommand{\gsim}{\lower.7ex\hbox{$\;\stackrel{\textstyle>}{\sim}\;$}}
\newcommand{\lsim}{\lower.7ex\hbox{$\;\stackrel{\textstyle<}{\sim}\;$}}
\newcommand{\TRH}{T_{\text{RH}}}
\newcommand{\CP}{\mathcal{P}}
\newcommand{\CR}{\mathcal{R}}
\newcommand{\C}{\mathbb{C}}
\def\unit{\relax{\rm 1\kern-.26em I}}
\def\OO{{\cal O}}
\def\CO{{\cal O}} 
\def\CL{{\cal L}}
\def\CM{{\cal M}}
\def\CO{{\cal O}}
\def\CL{{\cal L}}
\def\CM{{\cal M}}
\def\mpl{M_{\rm Pl}}
\newcommand{\bel}[1]{\be\label{#1}}
\def\al{\alpha}
\def\bt{\beta}
\def\eps{\epsilon}
\def\eg{{\it e.g.}}
\def\ie{{\it i.e.}}
\def\mn{{\mu\nu}}
\newcommand{\rep}[1]{{\bf #1}}
\def\be{\begin{equation}}
\def\ee{\end{equation}}
\def\bea{\begin{align}}
\def\eea{\end{align}}
\newcommand{\eref}[1]{(\ref{#1})}
\newcommand{\Eref}[1]{Eq.~(\ref{#1})}
\newcommand{\vev}[1]{ \left\langle {#1} \right\rangle }
\newcommand{\bra}[1]{ \langle {#1} | }
\newcommand{\ket}[1]{ | {#1} \rangle }
\newcommand{\ev}{{\rm eV}}
\newcommand{\kev}{{\rm keV}}
\newcommand{\Mev}{{\rm MeV}}
\newcommand{\gev}{{\rm GeV}}
\newcommand{\tev}{{\rm TeV}}
\newcommand{\mev}{{\rm MeV}}
\newcommand{\meV}{{\rm meV}}
\newcommand{\TeV}{\,\mathrm{TeV}}
\newcommand{\GeV}{\,\mathrm{GeV}}
\newcommand{\MeV}{\,\mathrm{MeV}}
\newcommand{\keV}{\,\mathrm{keV}}
\newcommand{\eV}{\,\mathrm{eV}}
\newcommand{\mnu}{\ensuremath{m_\nu}}
\newcommand{\nnu}{\ensuremath{n_\nu}}
\newcommand{\mlr}{\ensuremath{m_{lr}}}
\newcommand{\acc}{\ensuremath{{\cal A}}}

\newcommand{\fh}{\ensuremath{{\tilde h}}}
\newcommand{\fa}{\ensuremath{{\tilde a}}}
\newcommand{\fs}{\ensuremath{{\tilde s}}}
\newcommand{\h}{\ensuremath{{ h}}}
\newcommand{\A}{\ensuremath{{ a}}}
\newcommand{\s}{\ensuremath{{ s}}}
\newcommand{\Tr}{{\text{ Tr }}}
\newcommand{\tr}{{\text{ tr }}}

\title{Gravitational Waves from Supersymmetry Breaking}

\author{Nathaniel J. Craig}
\email{ncraig@stanford.edu}
\affiliation{Department of Physics, Stanford University, Stanford, CA 94305-4060}

\preprint{SU-ITP-09/08}

\begin{abstract}

In theories of supersymmetry breaking, it is often the case that there is more than one metastable vacuum. First-order phase transitions among such metastable vacua may generate a stochastic background of gravitational waves, the observation of which would provide a direct window into the supersymmetry-breaking sector.

\end{abstract}

\maketitle

The detection of gravitational waves (GWs) will provide unparalleled insight into the history of the early Universe. Uninfluenced by recombination, such gravitational waves shed light on phenomena of the earliest epochs. In particular, a stochastic background of GWs is expected to be produced by high-energy cosmological processes, including first-order phase transitions. 

Such first-order phase transitions are a common prediction of new physics. Electroweak symmetry breaking may entail a first-order phase transition at temperatures $T_* \sim 100$ GeV, while more speculative theories of new physics  -- such as grand unification or Randall-Sundrum models of warped extra dimensions -- may produce first-order phase transitions at temperatures ranging from $T_* \sim 1$ TeV to $T_* \sim 10^{16}$ GeV. 

Supersymmetry (SUSY), spontaneously broken at the electroweak scale, is among the most attractive candidates for new physics beyond the Standard Model. The breaking of supersymmetry typically occurs in a sector sequestered from the Standard Model, whose fields inhabit a supersymmetry-breaking vacuum. This vacuum need not be the global minimum of the theory; indeed, metastable supersymmetry-breaking appears to be a simple and generic feature of many SUSY-breaking sectors \cite{Dimopoulos:1997ww}. Such theories often feature one or more supersymmetry-breaking vacua with parametrically small rates for transition into supersymmetric vacua. However, it is quite possible for first-order phase transitions to occur rapidly among the supersymmetry-breaking vacua of the theory. 

In this paper, we wish to draw attention to the potential observation of gravitational waves from phase transitions among metastable supersymmetry-breaking vacua in the early Universe. Such a phase transition would be generically expected to occur at temperatures $T_* \sim \sqrt{F},$ where $\sqrt{F}$ is the primordial supersymmetry-breaking scale. A stochastic background of gravitational waves is generated when such a phase transition nears completion. When several bubbles of the final vacuum collide, the spherical symmetry of the bubbles is broken, and a fraction of the kinetic energy contained in the bubble walls is converted to gravitational waves. Moreover, the collision of two or more bubbles produces anisotropic stirring of the cosmic plasma, generating additional GWs. Gravitational waves from phase transitions of this genre may be measured at future ground- and space-based GW detectors, providing a direct window into the supersymmetry-breaking sector.

{\it Metastable Supersymmetry Breaking.--}If supersymmetry provides an explanation for the hierarchy problem, it must be broken in a sector separate from the fields of the Standard Model, most likely by some dynamical mechanism \cite{Witten:1981nf}. Indeed, supersymmetric gauge theories generically possess both supersymmetric and metastable supersymmetry-breaking vacua \cite{Intriligator:2006dd}. 

 In a typical theory with more than one metastable supersymmetry-breaking vacuum, it is generally the case that many features of the potential separating metastable vacua are set by one parameter: the primordial SUSY-breaking scale, $\sqrt{F}.$ In such theories, both the critical temperature $T_c$ at which first-order phase transitions commence, and the nucleation temperature $T_*$ at which the phase transition completes, are of order $\sim \sqrt{F}.$ Consequently, the spectrum of stochastic gravitational waves from a first-order phase transition among metastable supersymmetry-breaking vacua is characterized by $\sqrt{F}.$

	The size of the supersymmetry breaking scale $\sqrt{F}$ may be determined by the phenomenological requirement that the MSSM soft-scale masses lie around the weak scale, i.e., $F/M \sim 10^2 - 10^3$ GeV, where $M$ denotes the mass scale of the mechanism mediating supersymmetry breaking to the MSSM. In the case of gravity mediation, $M = M_P,$ which implies a primordial supersymmetry-breaking scale $\sqrt{F}_{\text{grav}} \simeq 10^{11}$ GeV. In the case of gauge mediation, however, the messenger scale $M$ may be as low as $10^4$ GeV (satisfying experimental limits on sparticle masses) and as high as $10^{15}$ GeV (above which flavor-violating contributions from gravity mediation become significant). Consequently, the supersymmetry breaking scale for gauge mediation may range from $\sqrt{F}_{\text{gauge}} \simeq 10^4 - 10^9$ GeV. Within this range, low-scale gauge mediation (corresponding to $\sqrt{F} \simeq 10^4 - 10^6$ GeV) leads to a phenomenologically favorable spectrum of MSSM soft masses \cite{Dine:1993yw}. 
	
	There are numerous SUSY-breaking models exhibiting more than one metastable supersymmetry-breaking vacuum (e.g., \cite{Intriligator:2006dd, Intriligator:1996fk}). It is often the case that the supersymmetry-breaking vacua of such theories are parametrically close in field space, while the supersymmetric vacua are generated by nonperturbative effects and thus are sufficiently distant to make SUSY-breaking vacua long-lived. Moreover, for sufficiently high reheating temperatures, thermal effects generally select vacua in which supersymmetry is broken \cite{Craig:2006kx}. 
	
	For example, consider supersymmetric $SU(N_c)$ QCD with $N_f$ fundamental and antifundamental chiral superfields $Q,  \tilde Q$ and adjoint $X$ with polynomial superpotential, as studied in \cite{Kutasov:1995ve}. This theory is strongly coupled in the infrared, but possesses an IR-free dual provided $N_c / 2 \leq N_f < 2 N_c /3.$ The adjoint scalar potential breaks the gauge group into a `landscape' of vacua corresponding to the different vacuum expectation values of $X.$ In the presence of a mass deformation $m Q \tilde Q$ in the superpotential, the low-energy theory consists of many decoupled copies of the ISS model of metastable SUSY breaking \cite{ Intriligator:2006dd} with gauge groups of varying size. The supersymmetric vacua of the theory are parametrically distant from the supersymmetry-breaking vacua, and tunneling rates into supersymmetric vacua may naturally be made longer than the age of the universe. The supersymmetry-breaking vacua, however, are separated by barriers and distances of order $~ m_X \sim \sqrt{F},$ and tunneling rates between these vacua may be large. 
	
	Moreover, if the Universe is reheated to a temperature $T_{RH} \gtrsim \sqrt{F}$ after inflation, thermal effects will generically select the supersymmetry-breaking vacuum with the largest gauge group. For the $SU(N_c)$ SQCD theory with adjoints, in the infrared theory this corresponds to selecting the vacuum with the {\it largest} value of $\sqrt{F}.$ After the metastable vacuum with the largest scale of supersymmetry breaking is populated, transitions to metastable vacua with smaller $\sqrt{F}$ may rapidly take place \cite{NJC}. As such, in these theories it is reasonable to expect one or more first-order phase transitions among different metastable SUSY-breaking vacua in the early universe at temperatures $T_* \simeq \sqrt{F}$.\footnote{Whether these phase transitions give rise to topological defects depends on the particular dynamics of the supersymmetry-breaking sector, but the scales of interest are too low to produce a troublesome overdensity of defects.}
	
{\it Gravitational Waves.--}In general, a first-order phase transition will occur whenever the Universe initially finds itself in a false vacuum, separated from additional vacua of lower energy by a potential barrier. As temperature drops with the expansion of the universe, it becomes energetically favorable to nucleate bubbles of a lower-energy final vacuum within the initial vacuum phase. Bubbles above a critical size expand and collide, completing the transition into the final vacuum. The temperature $T_*$ at which the phase transition completes may be properly defined as the temperature at which the probability of nucleating one bubble per horizon volume, per horizon time, approaches unity. This condition amounts to the requirement $\Gamma(T_*)= H_*^4,$ where $\Gamma$ is the nucleation rate for bubbles of the final vacuum and $H_*$ is the Hubble parameter at temperature $T_*.$ 

As the phase transition comes to completion, collisions among bubbles of the final vacuum produce gravitational waves of characteristic frequency $f_*$ and fraction of the energy density $\Omega_{\text{gw} *}.$ These gravitational waves propagate to the present era without interacting, resulting in an observed frequency $f = f_* (a_*/a_0)$ and observed energy density fraction $\Omega_{\text{gw}} = \Omega_{\text{gw}*} (a_*/a_0)^4 (H_*/H_0)^2;$ $a(t)$ is the cosmological scale factor. The present contribution of gravitational waves to the energy density as a function of frequency is then
\beq
\Omega_{\text{gw}}(f) \equiv \frac{1}{\rho_c} \frac{d \rho_{\text{gw}}}{d \ln f}
\eeq
where $\rho_c$ is the critical density and $\rho_{\text{gw}}$ is the energy density in gravitational waves.

For sufficiently strong first-order phase transitions, the characteristic frequency $f_*$ and energy density of gravitational waves $\Omega_{\text{gw}*}$ produced at $T_*$ depends only weakly on the specific microphysics of the phase transition. Rather, it may be characterized by two parameters: $\alpha = \epsilon/\rho_{\text{rad}},$ the ratio between the latent heat liberated in the phase transition and the energy density in the high energy phase; and $\beta,$ the rate of time variation of the nucleation rate at transition temperature $T_*.$ The strength of the phase transition is controlled by $\alpha,$ with $\alpha \rightarrow 0$ and $\alpha \rightarrow \infty$ corresponding to weakly and strongly first-order phase transitions, respectively. The duration of the phase transition is given by $\beta^{-1}$ and the size of bubbles by $R_b \sim v_b /\beta,$ where $v_b$ is the velocity of the bubble wall. 

 The rate per unit volume for formation of bubbles of the final vacuum is $\Gamma \sim T^4 e^{-S},$ where $S(t)$ is the Euclidean bounce action. Correspondingly, $\beta$ is given in terms of the bounce action by $\beta = H_* T_* \left. \frac{dS}{dT} \right|_{T=T_*}.$ The percolation condition $\Gamma(T_*)= H_*^4$ then implies \cite{Kosowsky:1992rz}
\beq
\frac{\beta}{H_*} \sim 4 \ln \left( \frac{M_P}{T_*} \right).
\label{eqn:beta}
\eeq
Thus the quantity $H_*/ \beta,$ which sets the duration of the phase transition and typical bubble size, is expected to depend only logarithmically on the scale of the phase transition. For the thermal effective potentials of interest here, a strongly first -order phase transition $\alpha \gtrsim 1$ entails $T_* \simeq T_c / \sqrt{\text{few}},$ consistent with the transition temperatures implied by the percolation condition.

Loosely speaking, the characteristic frequency of GWs produced by bubble collisions is given by the time scale of the phase transition, $f_* \simeq \beta.$ Similarly, scaling arguments alone suggest that $\Omega_{\text{gw}*}$ is given in terms of $\beta/H_*$ by \cite{Grojean:2006bp}
\beq
\Omega_{\text{gw}*} = \frac{\rho_{\text{gw}*}}{\rho_{\text{tot}*}} \sim \kappa^2 v_b^3 \frac{\alpha^2}{(1+\alpha)^2} \left[ \frac{H_*}{\beta} \right]^2
\eeq
where $\kappa$ is the fraction of vacuum energy transformed into fluid kinetic energy. The spectrum of gravitational waves produced by subsequent turbulent motions of the plasma is somewhat more complicated, and depends additionally on the turbulent fluid velocity $u_s.$ In particular, the peak frequency of GWs produced by turbulence is set by the largest bubbles involved in the phase transition, and thus is lower than the peak frequency produced by collisions. 

For the case of detonation collisions, in which the bubble wall propagates faster than the speed of sound (as is the case for strongly first-order phase transitions), the bubble wall velocity $v_b,$ fluid velocity $u_s,$ and energy fraction $\kappa$ are given simply in terms of $\alpha$ by \cite{Steinhardt:1981ct,Kamionkowski:1993fg}
\beq
v_b(\alpha) &= &\frac{1/\sqrt{3} +\sqrt{\alpha^2+2\alpha/3}}{1+\alpha} \\
u_s(\alpha) &\simeq& \sqrt{\frac{\kappa \alpha}{\frac{4}{3}+\kappa \alpha}} \\
\kappa(\alpha) &\simeq& \frac{1}{1+0.715 \alpha}\left[0.715 \alpha +\frac{4}{27}\sqrt{\frac{3\alpha}{2}}\right].
\eeq
More precise estimates of the peak frequency and energy density stored in GWs due to bubble collisions have been made on the basis of numerical simulations  \cite{Kosowsky:1992rz,Kamionkowski:1993fg}. More recently, an accurate analytic approximation for the power spectrum was obtained for gravitational waves produced by bubble collisions, in which the present-era peak amplitude and frequency of GWs from bubble collisions are found to be \cite{Caprini:2007xq}
\beq
h^2 \Omega^c_{\text{peak}} \simeq 9.8 \times 10^{-8}  \frac{v_f^4 (1-s^3)^2}{(1-s^2 v_f^2)^4} 
 \left[\frac{{H}_*}{\beta}\right]^2 \left[\frac{100}{g_*}\right]^{\frac{1}{3}} \\
 f_{\text{p}}^c \simeq 1.12 \times 10^{-7}\mbox{ Hz} \  \frac{1}{v_b} \left[\frac{\beta}{H_*}\right] \left[\frac{T_*}{1 \mbox{GeV}}\right] \left[\frac{g_*}{100}\right]^{1/6}.
\eeq
Here  $s = \frac{1}{\sqrt{3} v_b};$ and $v_f = \frac{v_b - 1/\sqrt{3}}{1-v_b/\sqrt{3}}$ is the outer maximal fluid velocity. 

Similarly, the GW spectrum due to plasma turbulence has been the subject of extensive, albeit more recent, numerical study \cite{Caprini:2006jb}, according to which the peak amplitude and frequency for GWs produced by plasma turbulence are given by
\beq
h^2 \Omega^t_{\text{peak}}  \simeq 8.1 \times 10^{-7} v_b^2 u_s^2  \left[\frac{H_*}{\beta}\right]^2 \left[\frac{100}{g_*}\right]^{1/3}\\
f_{\text{p}}^t \simeq 8 \times 10^{-8} \mbox{Hz} \frac{1}{v_b} \left[\frac{\beta}{H_*}\right]\left[\frac{T_*}{1 \mbox{GeV}}\right]\left[\frac{g_*}{100}\right]^{1/6} 
\eeq
provided $u_s \geq 1/2$ (as is the case for $\alpha \gtrsim 0.9$); for $u_s < 1/2,$ the peak amplitude is reduced by a factor of $4 u_s^2.$

\begin{figure}[t] 
   \centering
   \includegraphics[width=3.3in]{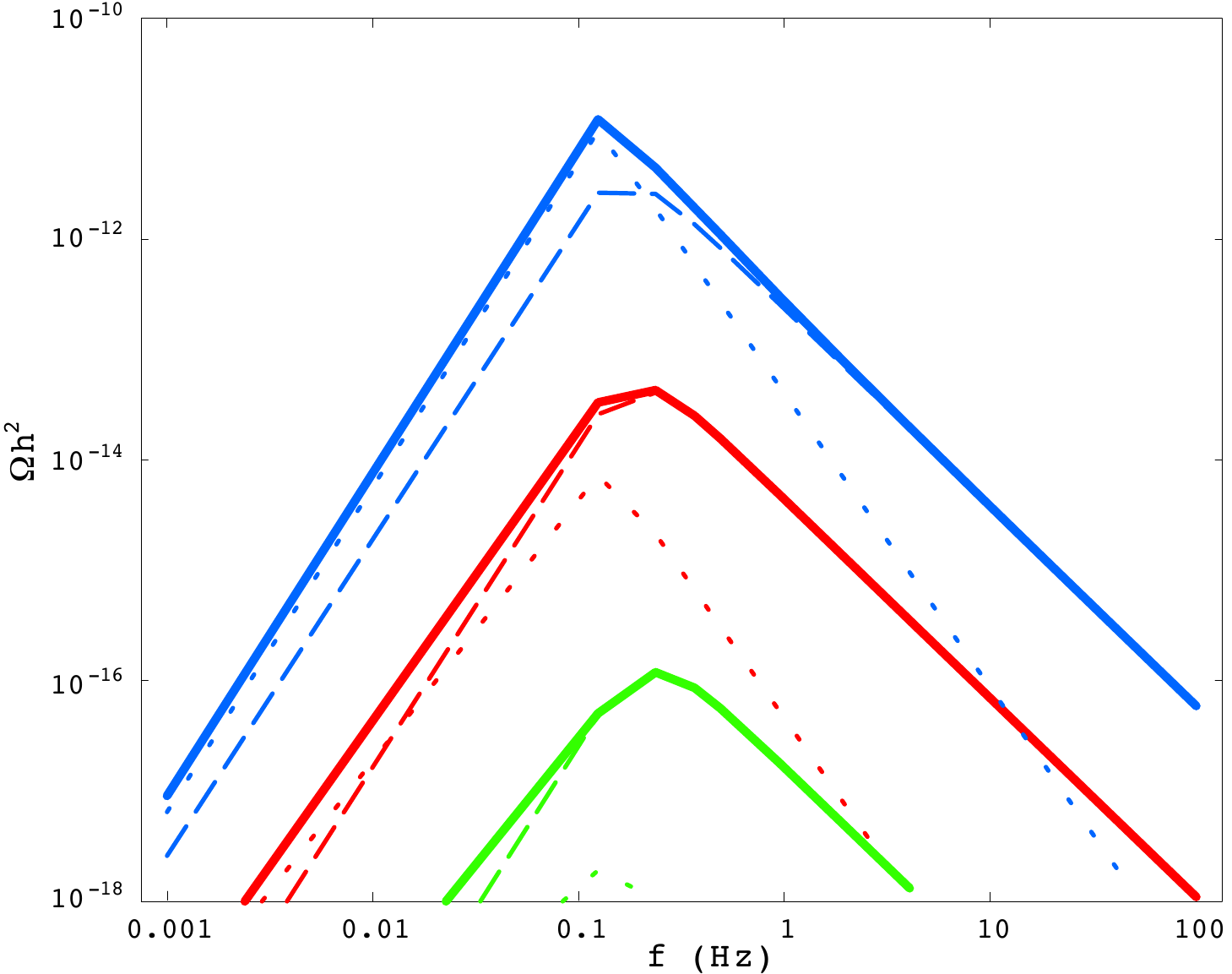} 
   \caption{Stochastic gravitational wave signal of a first-order supersymmetry-breaking phase transition with $\sqrt{F} = 10^4$ GeV for varying strengths of the phase transition: $\alpha = 0.01$ (green),  0.1 (red), and 1.0 (blue). Dotted lines denote the signal from turbulence; dashed lines denote the signal from collisions; the solid line is the total stochastic signal for each value of $\alpha$.}
   \label{fig:sig}
\end{figure}

Both collisions and turbulence may contribute significantly to the gravitational wave spectrum, with turbulence becoming dominant as the phase transition strengthens. The peak frequency of the resulting spectrum of gravitational waves directly reflects the temperature at which the phase transition occurs, while the energy density in gravitational waves depends on the strength of the phase transition. The above results, coupled with expressions for the full gravitational wave spectrum $\Omega_{\text{gw}}(f)$ for collisions \cite{Caprini:2007xq} and turbulence \cite{Caprini:2006jb}, allow us to estimate GW spectra for various SUSY-breaking phase transitions. Characteristic gravitational wave spectra for first-order phase transitions with $\sqrt{F} \sim 10^{4}$ GeV and various values of $\alpha$ are shown in Fig. \ref{fig:sig}. 

\begin{figure}[t] 
   \centering
   \includegraphics[width=3.5in]{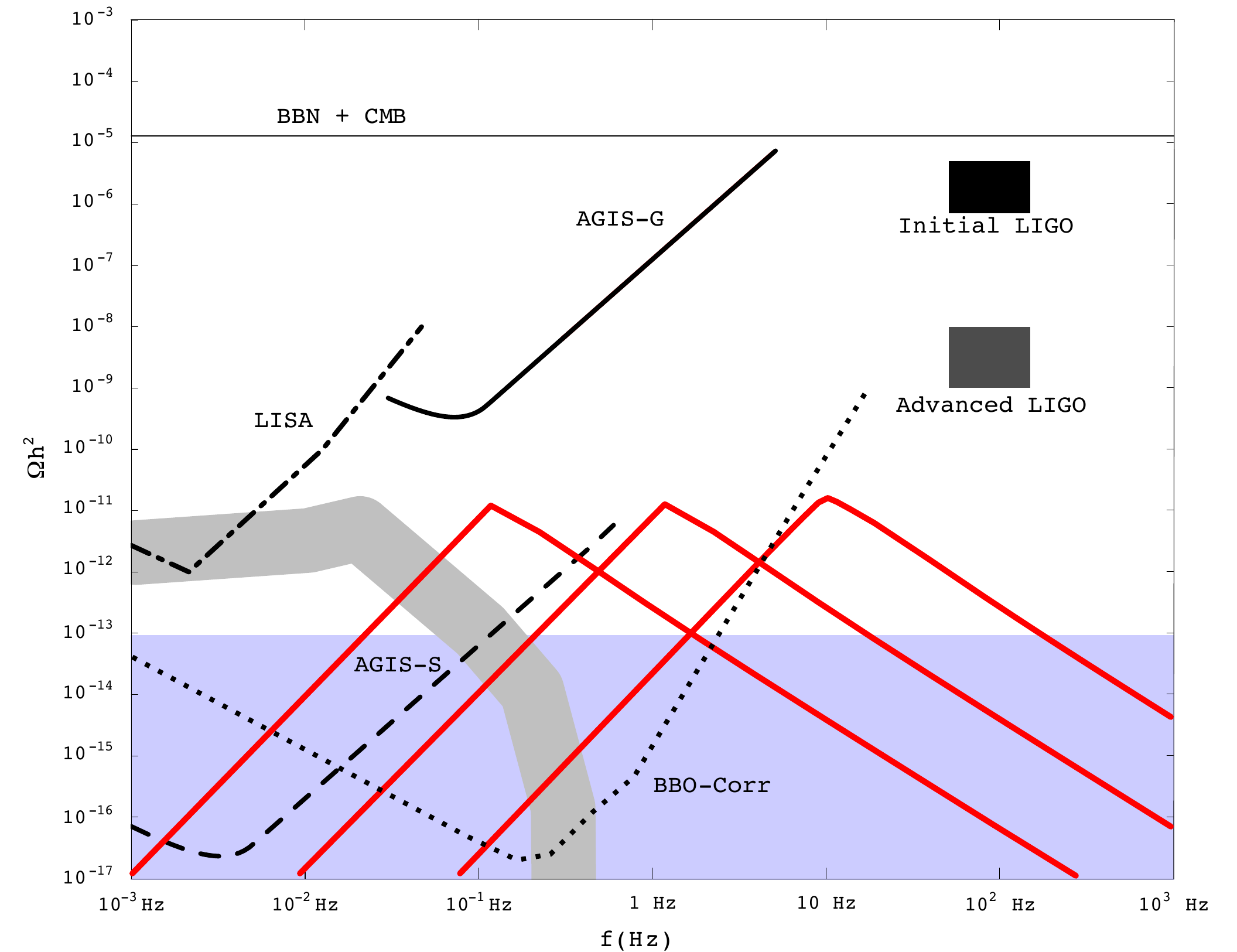} 
   \caption{GW spectra alongside sensitivity curves of various gravitational wave detectors. From left to right, the red peaks correspond to GW spectra with $\sqrt{F} = 10^4, 10^5, 10^6$ GeV and $\alpha = 1$. The solid black curve denotes the sensitivity of a ground-based AGIS experiment (AGIS-G); the dashed curve denotes that of the most aggressive space-based AGIS experiment (AGIS-S). The dashed-dotted curve denotes the sensitivity of LISA, and the dotted curve the sensitivity of a correlated BBO. The AGIS sensitivity curves are taken from \cite{Dimopoulos:2007cj}; the BBO sensitivity curves from \cite{Buonanno:2004tp}. The grey band shows the estimated background from extragalactic white dwarf binaries \cite{Farmer:2003pa}, while the blue band shows the region of anticipated signals from inflation.}
   \label{fig:sens}
\end{figure}

{\it Detection.--}The stochastic background of gravitational waves from SUSY-breaking phase transitions is likely too weak to be seen by ground-based bar detectors or  laser interferometers such as LIGO, VIRGO, GEO and TAMA. On the other hand, future ground-based experiments and proposed space-based detectors such as LISA, BBO, and AGIS \cite{Dimopoulos:2007cj} may be sufficiently sensitive to probe phase transitions with a low scale of supersymmetry breaking, as shown in Fig. \ref{fig:sens}. The peak sensitivity of LISA lies at a frequency well below the lowest frequencies expected from SUSY-breaking phase transitions. However, a spaced-based atom-interferometric experiment such as AGIS or a correlated laser experiment such as BBO would be sensitive to low-scale supersymmetry breaking with $\sqrt{F} \sim 10^{4} - 10^{6}$ GeV, while a future ground-based interferometer may probe $\sqrt{F} \sim 10^6 - 10^8$ GeV. Provided the first-order phase transition is sufficiently strong, the GW signal would be visible above anticipated stochastic backgrounds from astrophysical sources and inflation. Unfortunately, the gravitational waves produced in theories of gravity mediation (corresponding to peak frequencies $f_{\text{p}} \simeq$ 1 MHz) would generally lie at frequencies too high to be measured by space-borne experiments.

{\it Summary.--}Stochastic gravitational waves from cosmological processes in the early Universe provide unparalleled insight into the earliest epochs. If the sector responsible for weak-scale supersymmetry breaking possesses more than one metastable vacuum, first-order phase transitions among these supersymmetry-breaking vacua may generate a stochastic background of gravitational waves measurable by next-generation ground- and space-based interferometers. We expect a strongly first-order phase transition with supersymmetry-breaking scales $\sqrt{F} \sim 10^{4} - 10^{8}$ GeV, corresponding to low-scale gauge mediation, to be accessible by future ground- and space-based interferometers. The measurement of such gravitational waves would provide a direct window into the physics of the supersymmetry breaking sector.

{\it Acknowledgments.--}I would like to thank Savas Dimopoulos, Peter Graham, Daniel Green, Stuart Raby, and Surjeet Rajendran for useful comments and conversations. This work was supported by an NSF Graduate Research Fellowship, NSF contract PHY-9870115, and the Stanford Institute for Theoretical Physics.

\end{document}